\documentclass[letterpaper,10pt]{article}
\usepackage[T1]{fontenc}
\usepackage{parskip}
\usepackage[font={small}]{caption}
\usepackage[hidelinks]{hyperref}
\usepackage[margin=2.5cm]{geometry}
\usepackage{times}
\usepackage{amsmath,amssymb,amsfonts,amsthm,enumerate,multirow}
\usepackage[affil-it]{authblk}
\usepackage{listings}
\usepackage{graphicx} 
\usepackage{times} 
\usepackage{amsmath} 
\usepackage{amssymb}  
\usepackage{color}
\usepackage{caption}
\usepackage{subcaption}
\usepackage{siunitx}
\usepackage{hyperref}
\usepackage{floatrow}
\usepackage[section]{placeins}
\usepackage{etoolbox}
\usepackage{subcaption}

\usepackage{algorithm,algorithmic,url,listings}

\date{}

\newbool{changes}
\booltrue{changes}

\title{Bragg-Edge Elastic Strain Tomography for \textit{in situ} Systems from Energy-Resolved Neutron Transmission Imaging}
\author[1]{J.N. Hendriks\thanks{\url{Johannes.Hendriks@newcastle.edu.au}}}
\author[2]{A.W.T. Gregg}
\author[3]{C.M. Wensrich}
\author[4]{A.S. Tremsin}
\author[5]{T. Shinohara}
\author[6]{M. Meylan}
\author[7]{E.H. Kisi}
\author[8]{V. Luzin}
\author[9]{O. Kirsten}
\affil[1,2,3,7]{School of Engineering, University of Newcastle, Australia.}
\affil[4]{Space Sciences Laboratory, University of California, Berkeley CA 94720, USA}
\affil[5]{Materials and Life Sciences Facility, Japan Proton Accelerator Research Complex, Tokai-mura, Ibaraki 319-1195, Japan}
\affil[6]{School of Mathematical and Physical Sciences, The University of Newcastle, Australia}
\affil[8]{Bragg Institute, Australian Nuclear Science and Technology Organisation (ANSTO), Kirrawee NSW 2232, Australia}
\affil[9]{European Spallation Source, Lund 223 63, Sweden}

\begin{document}

\newcommand{\coverTitle}{Bragg-Edge Elastic Strain Tomography for \textit{in situ} Systems from Energy-Resolved Neutron Transmission Imaging}
\newcommand{\coverAuthors}{J.N. Hendriks, A.W.T. Gregg, C.M. Wensrich, A.S. Tremsin, T. Shinohara, M. Meylan, E.H. Kisi, V. Luzin, and O. Kirsten}
\newcommand{\coverStatus}{Published.}

\begin{titlepage}
    \begin{center}
        {\large \em Technical report}
        
        \vspace*{2.5cm}
        %
        {\Huge \bfseries \coverTitle  \\[0.4cm]}
        
        %
        {\Large \coverAuthors \\[2cm]}
        
        \renewcommand\labelitemi{\color{red}\large$\bullet$}
        \begin{itemize}
            \item {\Large \textbf{Please cite this version:}} \\[0.4cm]
            \large
            \coverAuthors. \coverTitle. \textit{Physical Review Materials}, 1.5 (2017): 053802 
        \end{itemize}
        
        \vfill

        \vfill
    \end{center}
\end{titlepage}

\maketitle

%
%
%

\begin{abstract}
Technological developments in high resolution time-of-flight neutron detectors have raised the prospect of tomographic reconstruction of elastic strain fields from Bragg-edge strain images.
This approach holds the potential to provide a unique window into the full triaxial stress field within solid samples.
While general tomographic reconstruction from these images has been shown to be ill-posed, an injective link between measurements and boundary deformations exists for systems subject to  \textit{in situ} applied loads in the absence of residual stress. 
Recent work has provided an algorithm to achieve tomographic reconstruction for this class of mechanical system.
This paper details an experiment focused on providing a proof-of-concept for this algorithm by carrying out a full tomographic reconstruction of a biaxial strain field within a non-trivial steel sample.
This work was carried out on the RADEN energy-resolved neutron imaging instrument at the Japan Proton Accelerator Research Complex with validation through Digital Image Correlation  and constant wavelength neutron strain scans from the KOWARI diffractometer within the Australian Centre for Neutron Scattering.
Aside from a small systematic error present within individual measurements, the experiment was a success, and now serves as a practical demonstration Bragg-edge transmission strain tomography for \textit{in situ} systems.
\end{abstract}



\section{INTRODUCTION}

Strain tomography has the potential to dramatically transform the way in which experimental mechanics is carried out. 
In much the same way that regular computed tomography has transformed medicine and other sciences, a full-field approach to measuring triaxial strain within solid materials will have an impact across a range of areas. 
For example, Additive Manufacturing (AM) processes allow tremendous flexibility in component design. 
AM components are now finding their way into critical aerospace applications, such as fuel nozzles \cite{tremsin2016investigation} and turbine blades \cite{watkins2013neutron}.
Residual strain locked in by the printing process has a critical impact on the strength of the resulting components and must be understood to advance this technology. 
In contrast to current destructive and semi-destructive techniques \cite{prime2001cross,standard2002standard}, 2D Digital Image Correlation (DIC) \cite{sutton2009image} and point-wise X-ray and neutron diffraction methods \cite{hauk1997structural,noyan2013residual,fitzpatrick2003analysis}, Bragg-edge neutron strain tomography promises to reconstruct the full three-dimensional strain field and hence stress within printed engineering parts on the scale of centimetres.

\begin{figure*}[!htbp]
    \centering
    \begin{subfigure}[b]{.6\linewidth}
    \includegraphics[width=1\linewidth]{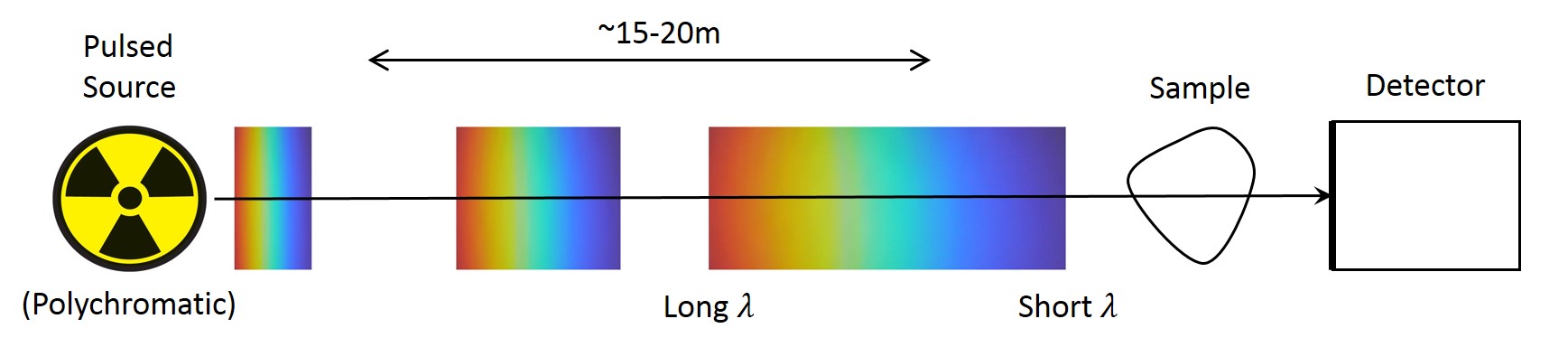}
    \caption{}
    \end{subfigure}
    \begin{subfigure}[b]{.2\linewidth}
    \includegraphics[width=1\linewidth]{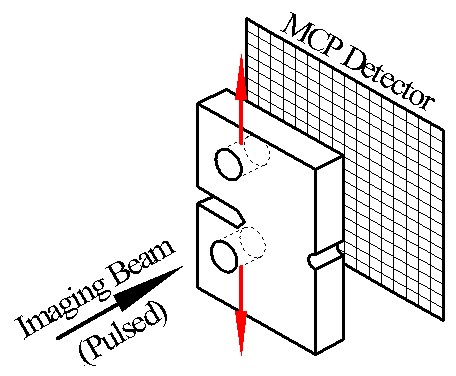}
    \caption{}
    \end{subfigure}
    \begin{subfigure}[b]{.45\linewidth}
    \includegraphics[width=1\linewidth]{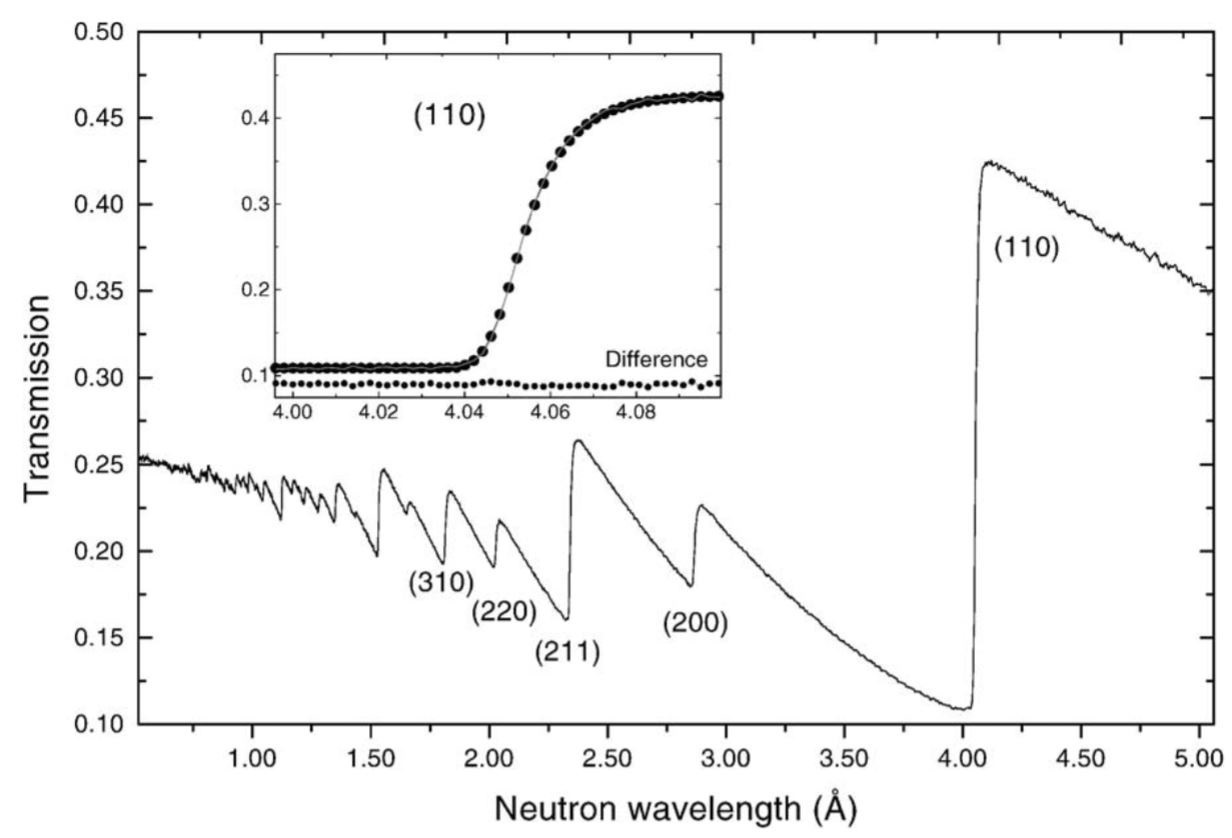}
    \caption{}
    \end{subfigure}
    \qquad
    \begin{subfigure}[b]{.47\linewidth}
    \includegraphics[width=1\linewidth,trim=0 -2.5cm 0 -2.5cm]{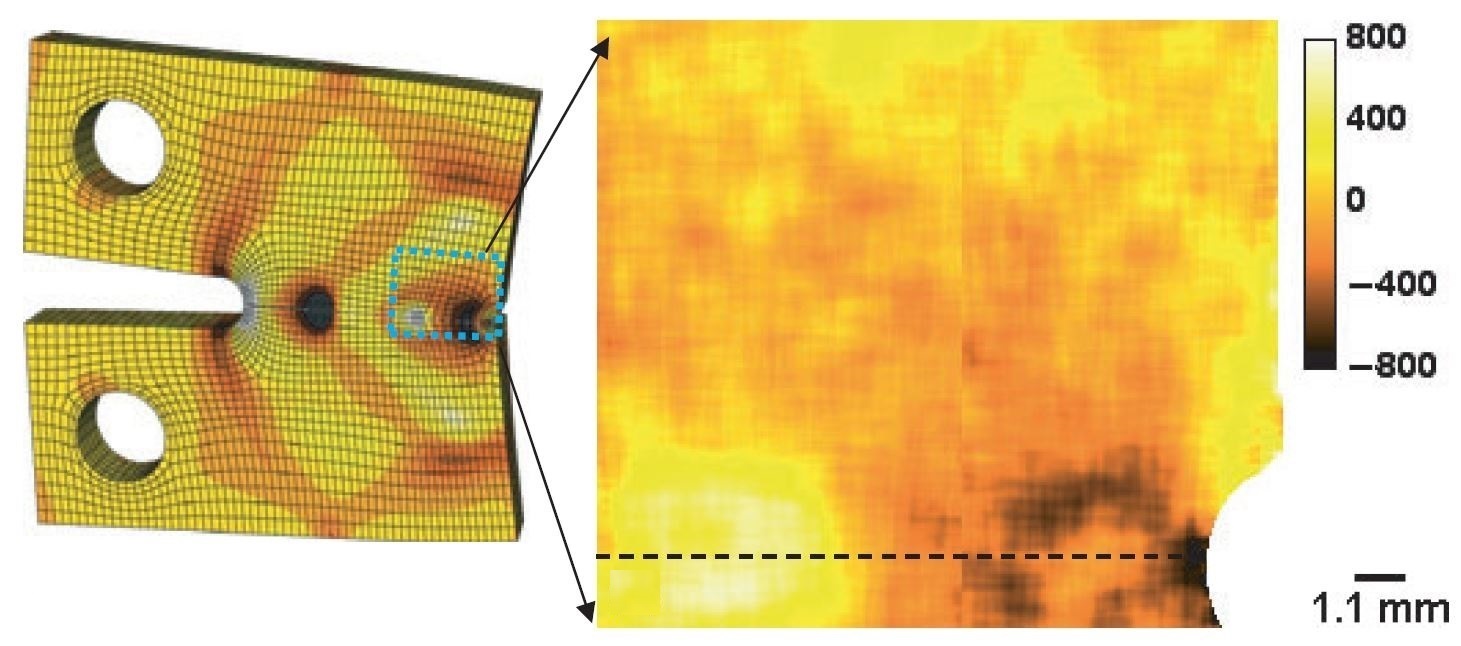}
    \caption{}
    \end{subfigure}
    \caption{(Colour online) (a) Energy-resolved Time-of-Flight (TOF) transmission geometry at a pulsed neutron source. (b) Bragg-edges in iron powder from a TOF transmission measurement \cite{santisteban2002engineering}. (c) and (d)  A high resolution strain image ($\times10^6$) from a two-dimensional steel compact-tension crack sample using a Micro-Channel Plate (MCP) detector by \cite{tremsin2012high} (Finite Element Model on the left, measurement on the right).}
    \label{fig:BraggEdges}
    \centering
\end{figure*}

The rapid development of pixelated `time-of-flight' detectors has made Bragg-edge analysis a prominent method for strain imaging.
Bragg-edges are sudden increases in the relative transmission through polycrystalline solids as a function of wavelength (see \autoref{fig:BraggEdges}b).

These edges are formed through diffraction; a neutron of wavelength $\lambda$ can be coherently scattered by crystal planes with spacing $d$, provided that the scattering angle, $\theta$, satisfies Bragg's law; $\lambda = 2d\sin\theta$. 
An abrupt increase in transmission occurs once $\lambda = 2d$ is exceeded as a neutron cannot be scattered by more than $2\theta = 180^\circ$ (i.e. backscattered). 
Like many diffraction techniques Bragg-edge location and geometry can provide a wealth of information on the internal structure of materials \cite{santisteban2002engineering}.

While other approaches are available (e.g. \cite{woracek2011neutron}), Bragg-edge analysis typically involves time-of-flight energy resolved measurements at pulsed neutron sources such as ISIS (UK), J-PARC (Japan) or SNS (USA). 
This technique relies upon the direct relationship between the velocity of a neutron and its wavelength. 
At a pulsed source, discrete bursts of polychromatic neutrons are created through spallation processes.
As each pulse travels down a flight tube (typically 15-$\si{20\meter}$) it becomes elongated as the fast neutrons (short wavelengths) outpace slower neutrons (long wavelengths) as illustrated in Figure \ref{fig:BraggEdges}a.
The time of arrival of each neutron is then a proxy for wavelength, allowing the relative transmission spectra through samples to be directly determined.

The position of a Bragg-edge can be obtained by fitting the integral form of the Kropff model \cite{kropff1982bragg} to the measured spectra. 
The relative shift in these positions can then provide a measure of strain of the form \cite{santisteban2002engineering};
\begin{equation}\label{eq:relative_shift}
    \langle \epsilon \rangle = \frac{\lambda-\lambda_0}{\lambda_0},
\end{equation}
where $\langle \epsilon \rangle$ is the measured normal strain in the beam direction, $\lambda$ is the position of the Bragg-edge, and $\lambda_0$ is the Bragg-edge position in an unstrained sample. 
Strains measured in this way are the average normal component in the direction of the transmitted ray. 
As with all diffraction techniques, these measurements refer only to the elastic component and hence are directly related to stress through Hooke's law.

 \begin{figure}[!htbp]
     \centering
     \begin{subfigure}[b]{.42\linewidth}
     \includegraphics[width=0.8\columnwidth]{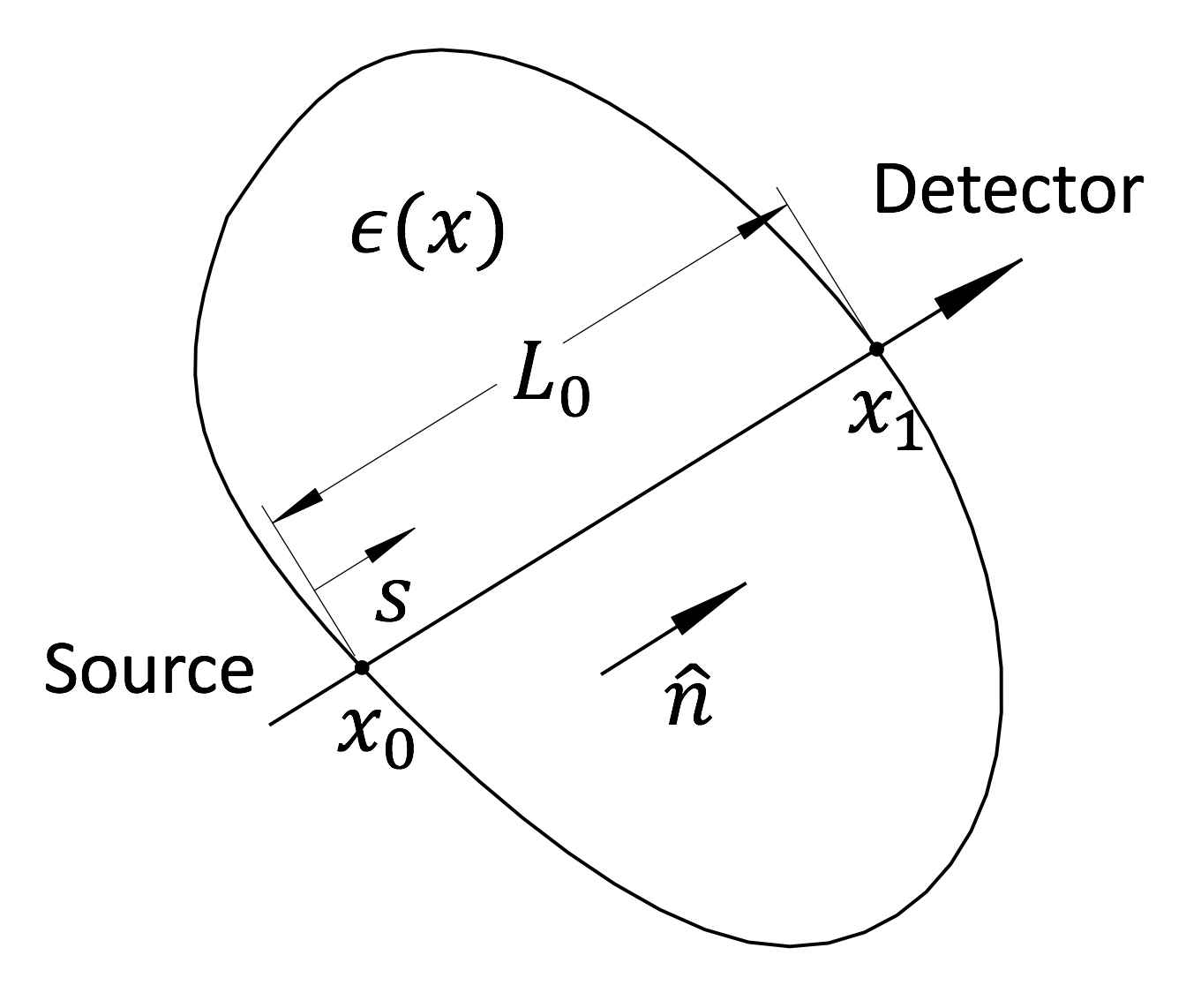}
     \caption{}
     \label{fig:2a}
     \end{subfigure}
     \qquad
     \begin{subfigure}[b]{.31\linewidth}
     \includegraphics[width=0.8\columnwidth]{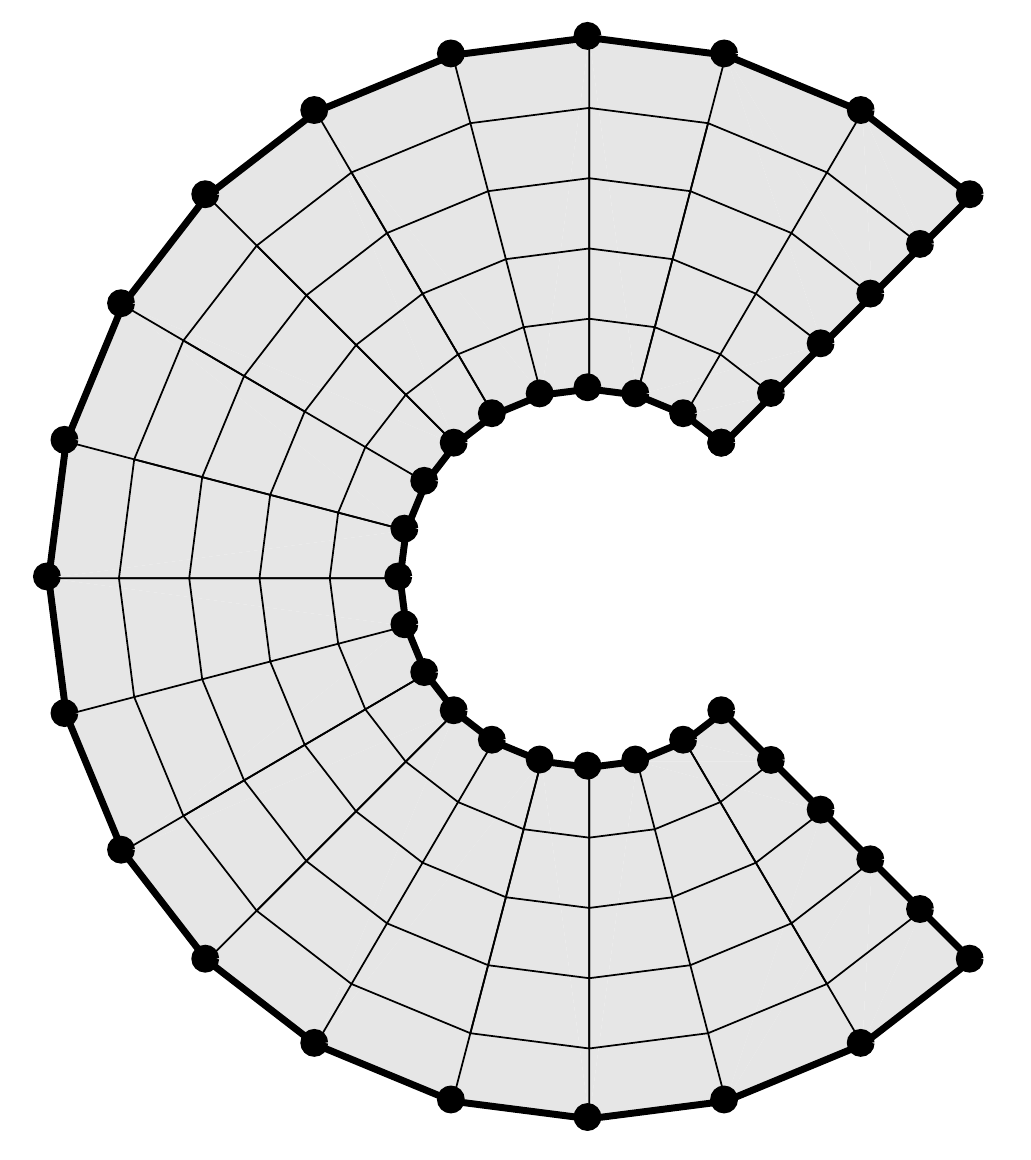}
     \caption{}
     \label{fig:2b}
     \end{subfigure}
     \caption{(a) The coordinate system for a single strain measurement modelled by the Longitudinal Ray Transform \eqref{eq:non_attenuated_integral}. (b) The boundary nodes and 2D Finite Element plate model used in the reconstruction process. }
     \label{fig:LRT}
 \end{figure}

Initially this work revolved around helium-based detectors \cite{santisteban2002strain}, however the current state-of-the-art $^{10}$B-based Micro-Channel Plate (MCP) technology due to Tremsin \textit{et al.} \cite{tremsin2011high} \cite{tremsin2009energy} is now capable of spatial resolutions down to $\si{55\micro\meter}$ \cite{tremsin2010transmission,tremsin2012high,strobl2012time} (e.g. see Figure~\ref{fig:BraggEdges}d). 

These developments have raised the tantalising prospect of strain tomography which has formed a focus for research over the past decade. 
As opposed to conventional tomographic imaging, this problem focuses on the reconstruction of a tensor field; a significantly more complex problem.

The measurement from a single ray observed by such a detector can be modelled by the Longitudinal Ray Transform (LRT);
\begin{equation}\label{eq:non_attenuated_integral}
    \langle \epsilon \rangle = \mathbf{I}_{\boldsymbol{\epsilon}} = \frac{1}{L_0}\int_{0}^{L_0} \epsilon_{ij}\left(\boldsymbol{x}_0+s\hat{\boldsymbol{n}}\right)\hat{n}_i\hat{n}_j \,\mathrm{d}s,
\end{equation}
where a ray in the direction $\hat{\boldsymbol{n}}$ enters the sample at $\boldsymbol{x}_0$, and $L_0$ is the irradiated sample length (see Figure~\ref{fig:2a}). Lionheart and Withers\cite{lionheart2015diffraction}, and prior work by Sharafutdinov \cite{sharafutdinov1994integral}, clearly demonstrates the general strain tomography problem from this transform is ill-posed. 
For conservative strain fields the LRT is only sensitive to boundary deformations, implying multiple strain fields can project to the same set of strain images (solutions to the problem are not unique). 
However, reconstruction can still be achieved if additional information is included through assumptions (special cases), or constraints imposed by the physics of the system.
For example, most of the initial work in this area is limited to special cases (e.g. axisymmetry \cite{abbey2009feasibility,abbey2012reconstruction,kirkwood2015neutron} and granular systems \cite{wensrich2016granular}).

A recent breakthrough in this area has provided a reconstruction technique for strain fields in samples subjected to \textit{in situ} loadings \cite{wensrich2016bragg}; i.e. boundary tractions in the absence of eigen-strain (e.g. plasticity), or body forces.
In this case, the lack of eigen-strain and body forces provides a direct link between the strain field and the boundary deformation \cite{wensrich2016bragg}.
This direct relationship provides a method for reconstruction.
Strain fields resulting from \textit{in situ} loadings represent a large class of mechanical systems; their reconstruction represents an important step towards general strain tomography.

This paper presents an experimental proof-of-concept for this algorithm; representing the first practical tomographic reconstruction of a non-axisymmetric strain field.

\section{EXPERIMENTAL SET-UP} 
\label{sec:experimental_set_up}

\begin{figure}[htbp]
    \centering
    \includegraphics[width=0.4\textwidth]{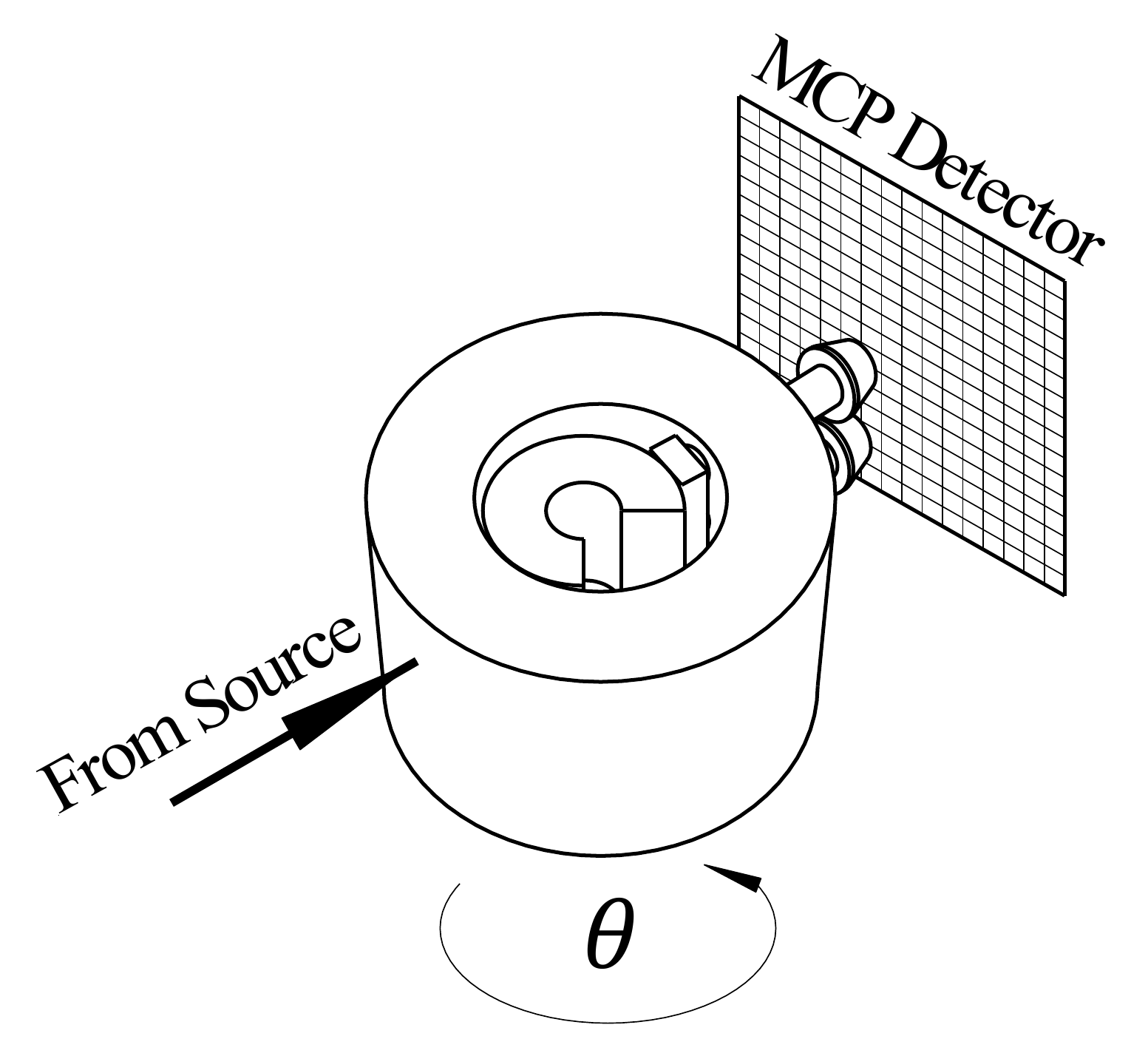}
    \caption{The experimental system and measurement geometry; a small, two-dimensional, C-shaped steel sample contained within an aluminium loading system positioned between the source and the MCP detector. The beam incident angle shown is for a stage rotation of $\theta=90^\circ$.}
    \label{fig:experiment_setup}
\end{figure}

This experiment was conducted over a 4 day period in November 2016 at the Japan Proton Accelerator Research Complex (J-PARC) on the RADEN energy resolved neutron imaging instrument. 
The system under examination consisted of a small, two-dimensional, C-shaped, EN-26 steel sample subject to a compressive load of approximately $\si{7\kilo\newton}$ (Figure~\ref{fig:experiment_setup}). 
The sample had an inner diameter of $\si{7\milli\meter}$, outer diameter of $\si{20\milli\meter}$ and thickness $\si{16\milli\meter}$.
Prior to loading, the sample was quenched and tempered at $560^\circ $C for one hour to relieve residual strains.
The resulting hardness of the material was around 250HV.
Load was applied by two M6 aluminium bolts (7075-T6) in a small die constructed from a high strength aluminium alloy (7075-T6). 
Aluminium was chosen to provide relative transparency to neutrons.
The two aluminium bolts were each torqued to $1.\si{7\newton\meter}$ to apply the $\si{7\kilo\newton}$ compressive load.

The sample geometry was carefully chosen to produce regions of high strain at some distance from the applied loads.
This creates a strain field that is trivial to model numerically; the effect of boundary conditions is limited to small regions around the contact points.

The sample and loading system were placed on a rotation stage between the beam source and an MCP detector \cite{tremsin2012high}, as illustrated in Figure~\ref{fig:experiment_setup}.
This detector consists of a $512\times 512$ array of pixels with a spatial resolution of \si{55\micro\meter} and nano second temporal resolution. 
The actual spatial resolution during the experiment was assessed using a `Siemens star' radial grating and found to be around \si{100\micro\meter}.

\begin{figure}[htbp]
    \centering
    \begin{subfigure}[b]{.4\linewidth}
    \includegraphics[width=1\textwidth]{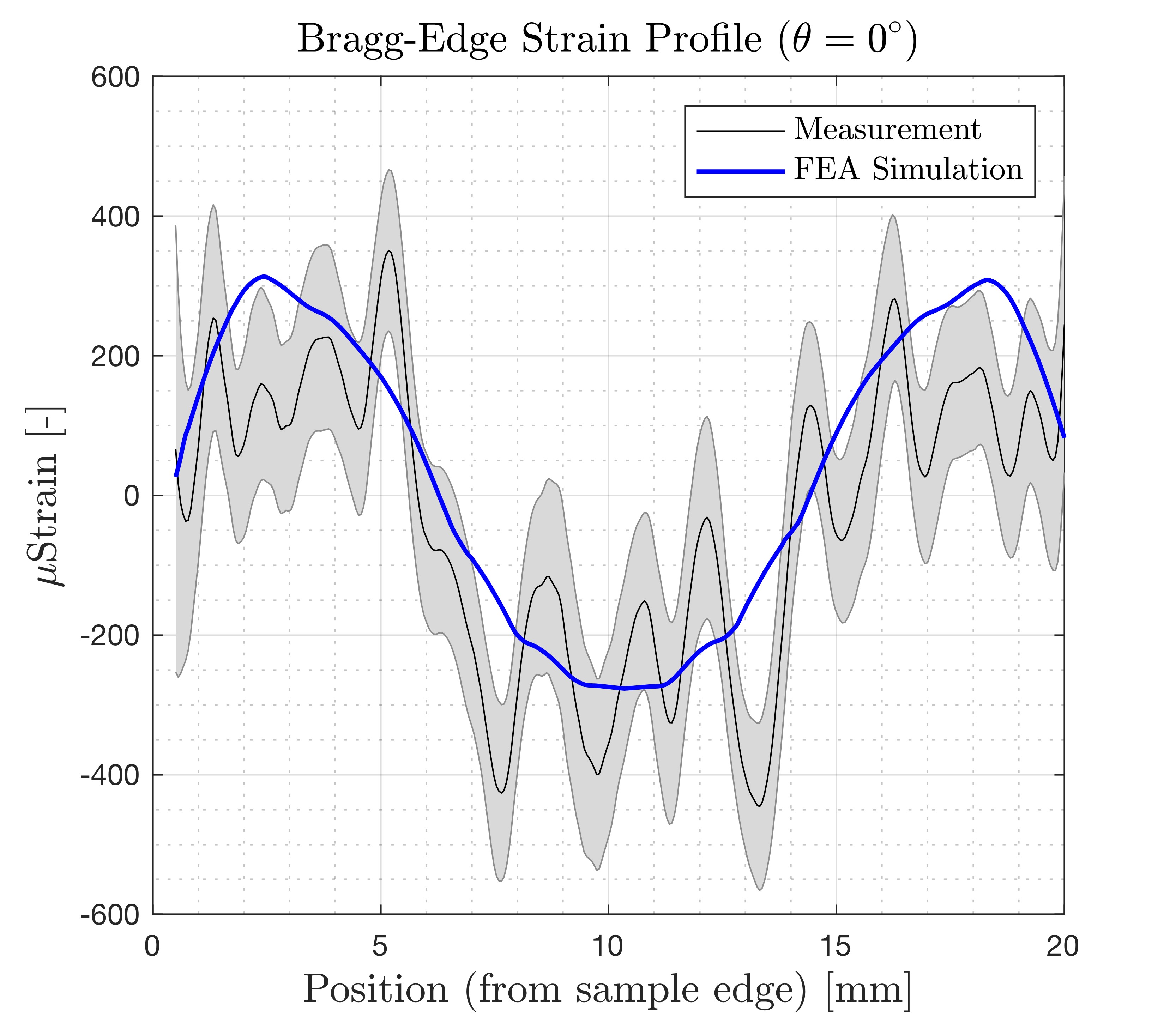}
    \caption{}
    \label{fig:a}
    \end{subfigure}
    \begin{subfigure}[b]{.4\linewidth}
    \includegraphics[width=1\textwidth]{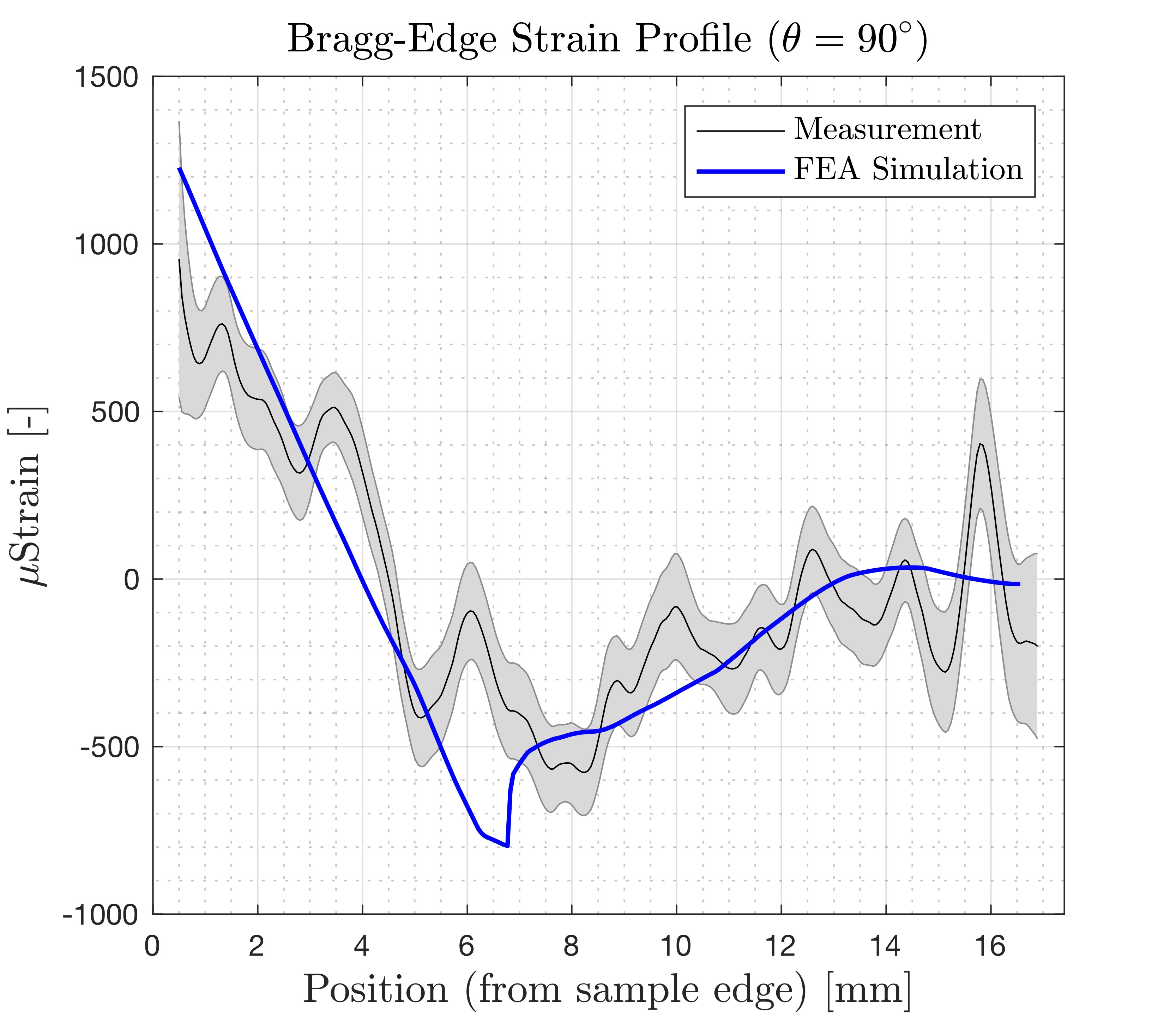}
    \caption{}
    \label{fig:a}
    \end{subfigure}
    \caption{A comparison between typical experimental strain profiles, and profiles generated from simulated FEA data. The grey band represents one standard deviation either side of the mean. (a) Profiles for $\theta=0^\circ$. (b) Profiles for $\theta=90^\circ$}
    \label{fig:strain_profile}
\end{figure}

After reference open-beam and $\lambda_0$ measurements, a series of 86 in-plane Bragg-edge strain projections were measured.
Each of these projections represents a one-dimensional strain profile through the sample in the transverse direction.
Figure~\ref{fig:strain_profile} shows two examples at different angles.
The projection angles, $\theta$, were based on golden angle increments to evenly distribute directions over $360^\circ$ irrespective of beam time interruptions.

Strain measurements were based on the relative shift of the $(110)$ Bragg-edge.
Each projection was sampled for one hour at a beam power of $\si{150\kilo\watt}$ and an $\si{18\meter}$ source distance.
A Gaussian image filter ($\sigma=4\text{ pixels}$) was applied to each time bin before grouping pixels into columns two pixels wide spanning the full sample length.
Given the constraints of available beam time, individual projections tended to be noisy. 
The grey band in Figure~\ref{fig:strain_profile} represents one standard deviation either side of the mean; this is a lower bound of the true uncertainty. 

Reconstruction from these profiles was achieved using the method presented by Wensrich \textit{et al.} \cite{wensrich2016bragg}. 
A summary of this approach is as follows;

Where the strains are a due to a continuous displacement field $\boldsymbol\phi$, the LRT \eqref{eq:non_attenuated_integral} can be written as;
\begin{equation}
    \mathbf{I}_{\boldsymbol{\epsilon}} = \frac{1}{L_0}(\phi_i\left(\boldsymbol{x}_0+L_0\hat{\boldsymbol{n}}\right) - \phi_i\left(\boldsymbol{x}_0\right))\hat{n}_i, 
\end{equation}
where $\phi_i\left(\boldsymbol{x}_0+L_0\hat{\boldsymbol{n}}\right)$ and $\phi_i\left(\boldsymbol{x}_0\right)$ are the boundary displacements at exit and entry locations of the ray. 
With the aid of linear shape functions, a system of equations relating $M$ measurements, $\langle \boldsymbol\epsilon \rangle$, to the displacements of $N$ discrete points along the boundary of the sample, $\boldsymbol\Phi$ can be written in the form;
\begin{equation}
    \mathbf\langle \boldsymbol\epsilon \mathbf\rangle_{[M\times1]}= \mathbf{A}_{[M\times2N]}\boldsymbol{\Phi}_{[2N\times1]}.
\end{equation}

A least squares approach can then be used to solve for $\boldsymbol\Phi$. 

From $\boldsymbol\Phi$, the internal strain field can then be found by solving the resulting Dirichlet boundary problem using Finite Element Analysis (FEA).
This method was slightly modified to account for rays passing through multiple segments in a non-convex sample, and was shown to converge to the true solution in simulation.

This algorithm requires detailed knowledge of the sample position, center of rotation, and orientation.
These were determined by matching the known sample geometry to edge height profile over all projections.

\section{RESULTS, VALIDATION AND ERROR ANALYSIS } 
\label{sec:results}
\begin{figure*}[!htb]
    \centering
    \includegraphics[width=0.9\textwidth]{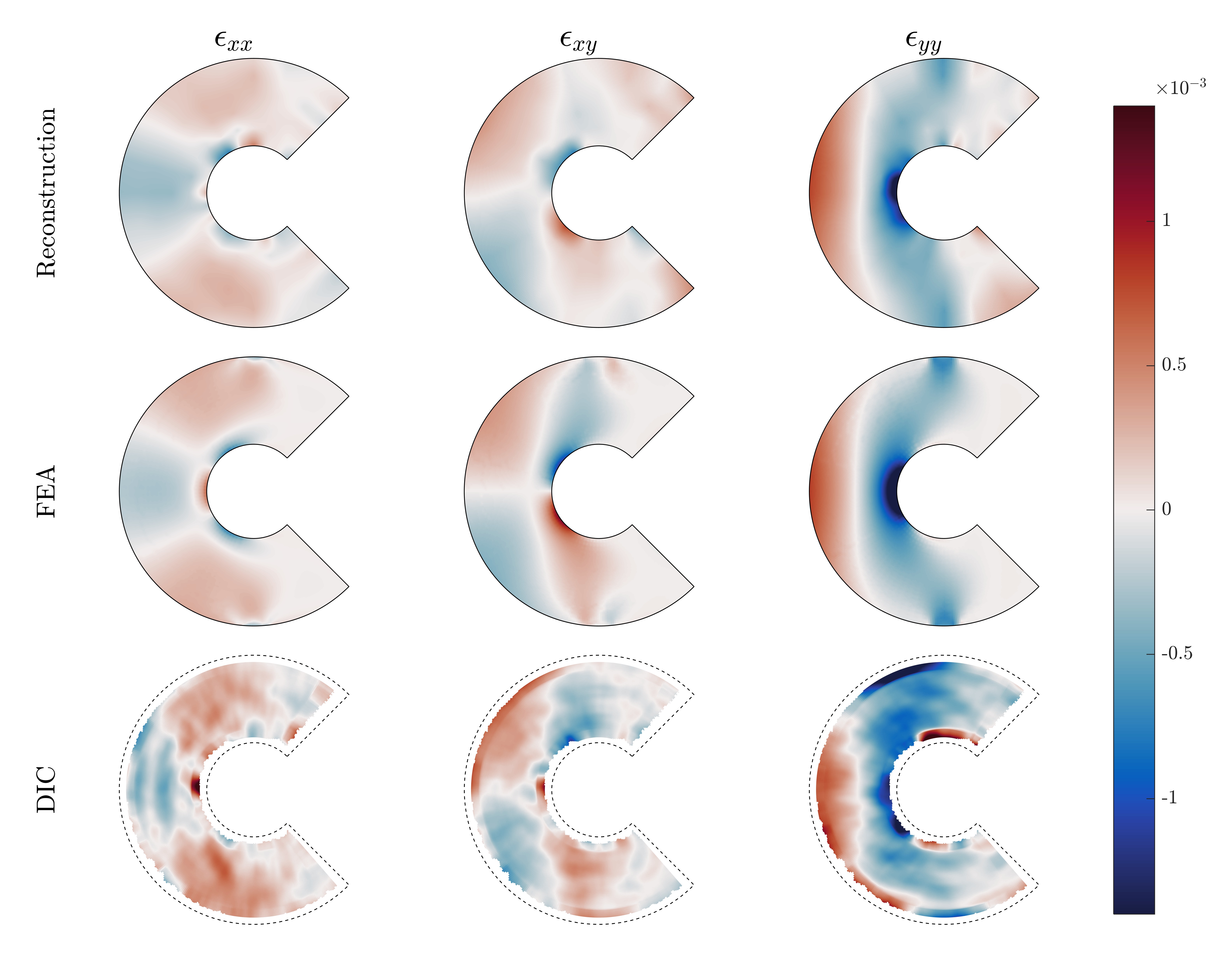}
    \caption{The reconstructed full-field two-dimensional strain components ($\epsilon_{xx}$, $\epsilon_{xy}$, and $\epsilon_{yy}$) compared to an FEA simulation and DIC results.}
    \label{fig:strain_comp}
\end{figure*}

The least-squares solution for 47 boundary nodes was found from the 86 projections, which then formed the boundary conditions for the two-dimensional, plane-stress, finite element model shown in  Figure \ref{fig:2b}).
Figure~\ref{fig:strain_comp} shows the reconstructed field in terms of the three cartesian components of strain; $\epsilon_{xx}$, $\epsilon_{xy}$, and $\epsilon_{yy}$. 
For $\epsilon_{yy}$, a band of compression is observed along the vertical axis in line with the applied load, with a maximum corresponding to $\sigma_{yy} = \si{-390\mega\pascal}$ concentrated on the inside surface.
A corresponding region of tensile strain exists on the left hand side as a result of the bending load (maximum of $\sigma_{yy}=\si{136\mega\pascal}$).
$\epsilon_{xx}$ and $\epsilon_{xy}$ have a less-direct relationship to the applied load.

\begin{figure*}[!htb]
    \centering
    \includegraphics[width=0.9\textwidth]{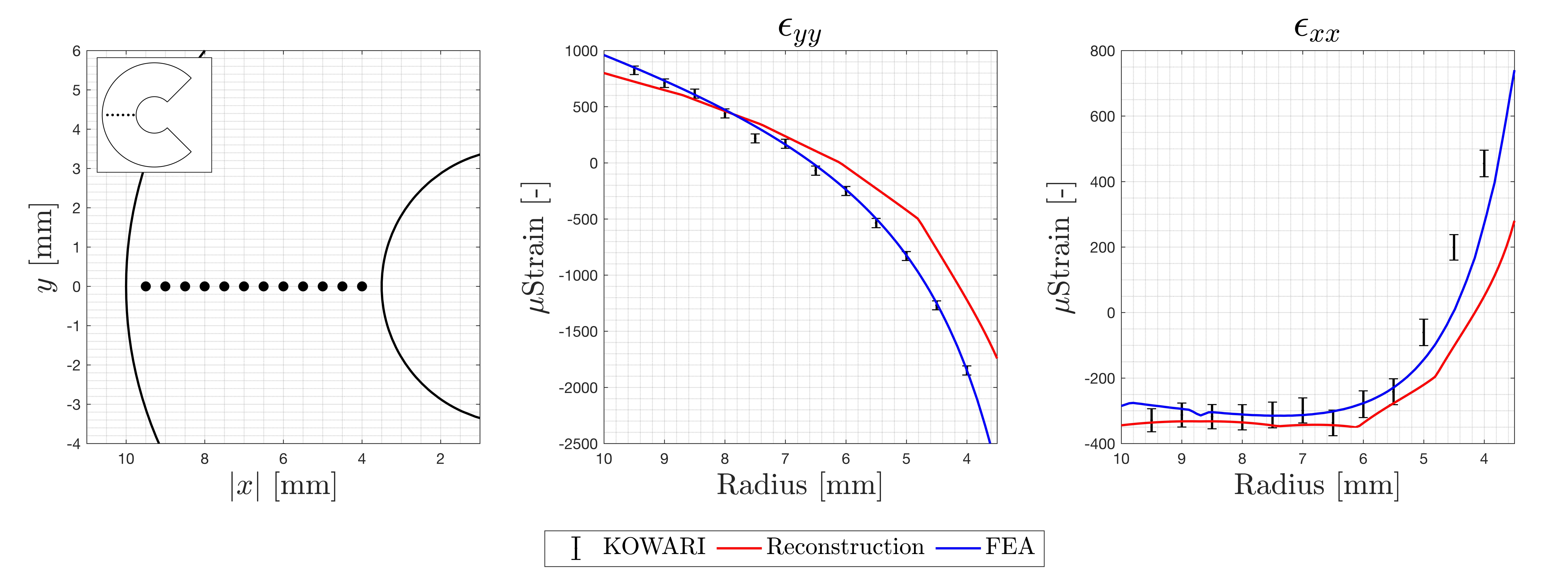}
    \caption{Comparison of the reconstructed and FEA strain fields to a constant wavelength strain scan from the KOWARI diffractometer.}
    \label{fig:KOWARI}
\end{figure*}

Figure~\ref{fig:strain_comp} also shows a corresponding strain field predicted from an FEA model.
Several assumptions are made by this model (e.g. symmetric uniform edge load, and plane stress), however it was validated was against constant wavelength neutron strain scanning on the KOWARI diffractometer at the Australian Center for Neutron Scattering within the Australian Nuclear Science and Technology Organisation (ANSTO).
This instrument provided measurements of $\epsilon_{xx}$ and $\epsilon_{yy}$ along the sample centreline based on the relative shift of the (211) $\alpha-\text{iron}$ diffraction peak using a gauge volume of $\si{1\times 1\times 14\milli\meter}^3$, and a wavelength of $1.67$\AA (i.e.  $90^\circ$ geometry).
Figure~\ref{fig:KOWARI} shows these measurements directly compared to the tomographic reconstruction and FEA results at corresponding locations.
The similarity between FEA predictions and the strain scan demonstrates the models suitability as a reference by which the quality of the reconstruction can be assessed.

DIC was also carried out during sample loading, the results of which are also shown in Figure~\ref{fig:strain_comp}.
DIC provides an estimate of full-field two-dimensional strain at the surface, by comparing digital images of the sample before and after loading \cite{pan2009two}. 
These strains are computed as the gradient of displacement field calculated through cross-correlation. 
Strain within most of the sample was at the lower end of what we could resolve using this technique, however orders of magnitude and structure were similar to the FEA model.

A comparison of the reconstruction to the FEA model, DIC results, and the KOWARI strain scan shows close agreement, however, the reconstruction tends to have a lower magnitude -- by approximately 15\% overall. 
Outside of this reduction in magnitude there are small regions of significant discrepancy near the inside surface of the `C'.
These discrepancies were greater than expected. 
This expected error was based on a set of simulations, that provide an estimate of the uncertainty from Gaussian measurement noise with standard deviation  $1\times 10^{-4}$.
This suggested convergence to the true solution with a standard deviation less than $5\times 10^{-5}$.

The reduction in magnitude was also observed in individual strain profiles (e.g Figure~\ref{fig:strain_profile}).
Similarly, discrepancies were observed in the profiles in areas corresponding to the inside surface of the `C'; for example the region from $5-7.\si{5\milli\meter}$ in Figure~\ref{fig:strain_profile}a.
This would suggests a systematic error is present in the strain measurement. 

This systematic error may arise from a variety of sources.
Of particular note, the regions of worst discrepancy tended to coincide with regions of high strain gradient.
The effect of strain gradients on the geometry of Bragg-edges is not well known.
The Kropff model \cite{kropff1982bragg,santisteban2002engineering} defines the geometry via five parameters, one of which specifies location.
This is assumed to have an independent and  direct relationship with average strain along the path; this may not be true when strain gradients exist. 

Another potential source of error is related to the spatial smoothing of required to reduce measurement uncertainty; this has the effect of tempering variation in the resulting profiles.
Longer sampling times or brighter sources may reduce the need for smoothing in future.
Note that beam divergence may play a similar role. 
It is important that work continues on resolving these issues.

\section{CONCLUSION} 
\label{sec:conclusion}

Aside from these systematic errors, this approach has performed well at reconstructing this strain field from experimental data. 
This represents the first tomographic reconstruction of a non-axisymmetric strain field from Bragg-edge measurements. 
This field was developed by an \textit{in situ} loading and the reconstruction hinged upon this assumption, however this special case is a significant step towards achieving general strain field tomography.

This demonstration provides a basis for future efforts focussed on extending to three-dimensional strain fields, as well as other approaches that may allow for general reconstruction of a broader class of strain field (e.g. residual strain).
A central component of this future work revolves around the development of algorithms suitable for the general case where plastic strains may present.
In this case, the link between internal strain and boundary deformation is severed, violating the key assumption of the algorithm demonstrated in this paper.
In place of this, future algorithms will be developed through a consideration of the physics of equilibrium or minimisation of strain energy to constrain the reconstructions to the true solution.
While these future algorithms continue to be a work-in-progress, the present work serves to provide a demonstration that the Bragg-edge transmission strain tomography has practical potential.

\section{ACKNOWLEDGEMENTS} 
\label{sec:acknowledgements}

This work is supported by the Australian Research Council through the Discovery Project scheme (ARCDP170102324). 
Access to RADEN instrument (J-PARC) and KOWARI (ANSTO) was made possible through their respective user programs (proposals 2016A0032 and P5293).
Additional support during the ANSTO experiment was provided by the Australian Institute for Nuclear Science and Engineering (AINSE).

\bibliographystyle{ieeetr}
\bibliography{References}

\end{document}